\documentclass[aps,pre,twocolumn,groupedaddress]{revtex4-1}
\usepackage[utf8]{inputenc}
\usepackage{amsmath}
\usepackage{amsfonts}
\usepackage{amssymb}
\usepackage{graphicx}
\usepackage{textcomp}
\DeclareMathSizes{10}{10}{6}{5}

\begin{document}

\title{Thermal conductivity of supercooled water}

\author{John W. Biddle}
\author{Vincent Holten}
\author{Jan V. Sengers}
\author{Mikhail A. Anisimov}\email{anisimov@umd.edu}
\affiliation{Institute for Physical Science and Technology, and Department of Chemical and Biomolecular Engineering, University of Maryland, College Park, Maryland 20742, USA}

\date{March 28, 2013}

\begin{abstract}
The heat capacity of supercooled water, measured down to $-37$ \textcelsius , shows an anomalous increase as temperature decreases. The thermal diffusivity, \emph{i.\hspace{1 pt}e.}, the ratio of the thermal conductivity and the heat capacity per unit volume, shows a decrease. These anomalies may be associated with a hypothesized liquid-liquid critical point in supercooled water below the line of homogeneous nucleation.  However, while the thermal conductivity is known to diverge at the vapor-liquid critical point due to critical density fluctuations, the thermal conductivity of supercooled water, calculated as the product of thermal diffusivity and heat capacity, does not show any sign of such an anomaly. We have used mode-coupling theory to investigate the possible effect of critical fluctuations on the thermal conductivity of supercooled water, and found that indeed any critical thermal-conductivity enhancement would be too small to be measurable at experimentally accessible temperatures. Moreover, the behavior of thermal conductivity can be explained by the observed anomalies of the thermodynamic properties.  In particular, we show that thermal conductivity should go through a minimum when temperature is decreased, as Kumar and Stanley observed in the TIP5P model of water. We discuss physical reasons for the striking difference between the behavior of thermal conductivity in water near the vapor-liquid and liquid-liquid critical points.
\end{abstract}

\maketitle

\section{Introduction}

Supercooled water exhibits several thermodynamic anomalies, perhaps the best known of which is its density maximum just above the freezing point, at 4~\textcelsius\ \cite{Debenedetti_2003b}.  The isobaric heat capacity \cite{Angell_1973,Archer_2000}, isothermal compressibility \cite{Speedy_1976}, and thermal expansivity \cite{Debenedetti_2003b} show anomalies that resemble critical-point power laws \cite{Speedy_1976}.  Moreover, the correlation length characterizing density fluctuations increases markedly upon supercooling \cite{Huang_2010}.  

In 1992, Poole \emph{et al.}\ proposed a coherent and particularly fruitful account of the thermodynamic anomalies of water: that deep in the supercooled region is a first-order phase transition between two liquid states distinguished by their different densities.  This transition line would terminate at a critical point, and proximity to this critical point could explain the anomalous behavior of the response functions \cite{Poole_1992}.  The liquid-liquid phase transition and critical point are hypothesized to occur at temperatures and pressures that are inaccessible to experiment due to unavoidable homogeneous nucleation of ice I$_\textrm{h}$ \cite{Debenedetti_2003b}.

This conjecture has given rise to several models, most of which propose two different ways in which water molecules might form hydrogen bonds.  One of these super-molecular arrangements (the high-density liquid or HDL) is associated with higher density and is favored at higher temperatures and higher pressures; the other (the low-density liquid or LDL) is associated with lower density and is favored at lower temperatures and lower pressures.

Intriguing experiments on the melting lines of metastable phases of D$_2$O and H$_2$O ice \cite{Mishima_1998,Mishima_2010} have lent additional support to this hypotheses, as has the finding of a phase transition in glassy water between two amorphous states of different density \cite{Mishima_1985}.  A recent study comparing the locus of glass-transition temperatures as a function of pressure in simulations of water molecules and in real water has found the behavior of real water to bear a stronger resemblance to the model that exhibits a liquid-liquid phase transition \cite{Giovambattista_2012}.  Other simulations \cite{Moore_2009,Kumar_2011,Cuthbertson_2011} note an increase in tetrahedrality in water upon supercooling, giving some idea as to the difference between the HDL and LDL states.  Equations of state based on the hypothesized existence of a liquid-liquid critical point have provided increasingly accurate accounts of the thermodynamic properties of water over the last several years \cite{Fuentevilla_2006, Bertrand_2011, Holten_2012a}.  The most recent two-state model of Holten and Anisimov (with entropy-driven liquid-liquid separation) \cite{Holten_2012b} shows excellent agreement with the thermodynamic data with fewer adjustable parameters than any model so far.  We have used the predictions of this two-state equation of state (which we abbreviate ``TSEOS") for thermodynamic properties in our calculations of the thermal conductivity.

While most of the phenomenology surrounding two-state models generally and the liquid-liquid critical point in particular has focused on thermodynamics, there are important implications for dynamics as well.  To take one example, the viscosity of water decreases upon compression, which is anomalous.  A suggested explanation for this anomaly is that compression forces a greater fraction of water into the HDL state, which has greater mobility than the tetrahedrally ordered LDL state \cite{Debenedetti_2003a}.  Furthermore, the dispersion of sound at high frequencies seems likely to reflect viscoelastic behavior associated with a structural relaxation in supercooled water \cite{Mallamace_2013}. 

In this paper we examine the thermal conductivity of supercooled water in light of the two-state conjecture, and we examine the possible effects that critical fluctuations might have on thermal conductivity.  According to our calculations, any effect of critical fluctuations associated with the virtual liquid-liquid critical point on the thermal conductivity would be too small to be measured at experimentally accessible temperatures.  Remarkably, the behavior of thermal conductivity can be fully explained by the anomalies of the thermodynamic properties.  The difference between the behavior of the thermal conductivity in the vapor-liquid and in the hypothesized liquid-liquid critical regions is the result of differences in both the dynamic and thermodynamic environments in the respective regions.

\section{Thermal Diffusivity and Thermal Conductivity}

The thermal diffusivity $a$ of water has been measured by Taschin \emph{et al.}\ down to 256~K \cite{Taschin_2006} and by Benchikh \emph{et al.}\ down to 250~K \cite{Benchikh_1985}, and it decreases steadily with decreasing temperature, as shown in Fig. 1.  Thermal conductivity $\lambda$ can be calculated from these data by means of the formula $\lambda = \rho c_p a$, since isobaric specific heat capacity $c_p$ and mass density $\rho$ are both known experimentally in the relevant temperature range.  The heat capacity changes little in the range for which thermal-diffusivity data are available (Fig. 2), so in that temperature range the thermal diffusivity and thermal conductivity are nearly proportional. At atmospheric pressure, thermal conductivity decreases with temperature, from the boiling point to the lowest temperatures at which thermal conductivity has been measured \cite{IAPWS_ThermalConductivity_2012}.  Water's behavior in this regard is unique among non-metallic liquids of low molecular weight, as all other such liquids show an increase in thermal conductivity upon cooling \cite{Adrianova_1967}.

\begin{figure}[h!]
\includegraphics[width=0.49\textwidth]{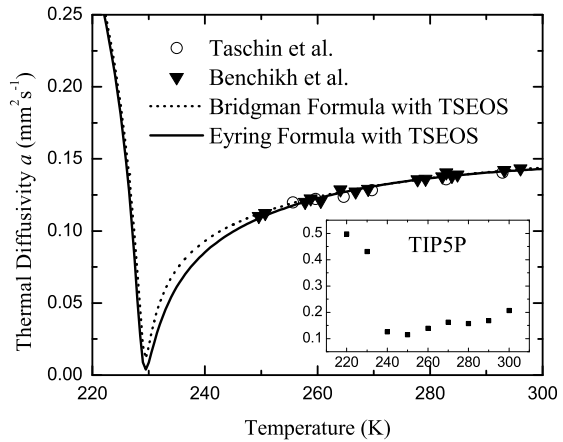}
\caption{Thermal-diffusivity data of Benchikh \emph{et al.}\ \cite{Benchikh_1985} and Taschin \emph{et al.}\ \cite{Taschin_2006}, compared with the thermal diffusivity calculated from Bridgman's (4) and Eyring's (5) formulae.  We use the TSEOS to evaluate the thermodynamic properties in these formulae \cite{Holten_2012b}.  The inset shows simulation results of Kumar and Stanley for the TIP5P model \cite{Kumar_2011}.}
\end{figure}

\begin{figure}[h!]
\includegraphics[width=0.49\textwidth]{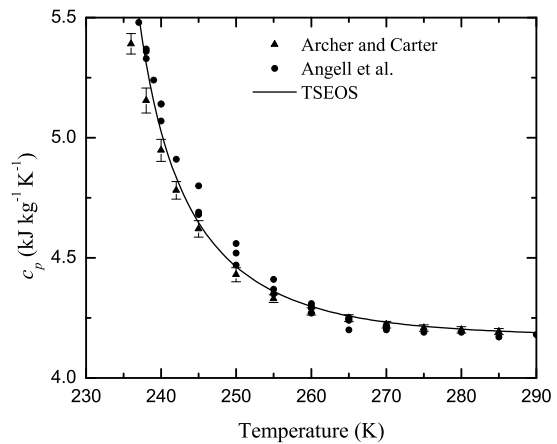}
\caption{Heat capacity data of Archer and Carter \cite{Archer_2000} and Angell \emph{et al.} \cite{Angell_1982}.  The solid curve shows the prediction of the TSEOS \cite{Holten_2012b}.}
\end{figure}

Benchikh \emph{et al.} have noted a strong correlation between the thermal conductivity of both supercooled and stable liquid water at low temperatures and the thermodynamic speed of sound $c$ (in the limit of zero frequency)  \cite{Benchikh_1985}.  We find excellent agreement between the experimental data and two classical formulations for the thermal conductivity of liquids, as shown in Fig. 3.  The first is Bridgman's formula \cite{Bridgman_1923} as adapted for polyatomic molecules:
\begin{equation}
\lambda = 2.8k_{\textrm{B}}v^{-2/3}c,
\end{equation}
where $k_\textrm{B}$ is Boltzmann's constant and $v$ is the molecular volume of the liquid, that is, the molecular mass divided by the mass density $\rho$.  The second formulation is due to Eyring, with a correction from Eucken \cite{Hirschfelder_1954}:
\begin{equation}
\lambda = 2.8k_{\textrm{B}}v^{-2/3}\gamma^{-1/2}c,
\end{equation}
where $\gamma$ is the ratio of the isobaric heat capacity to the isochoric heat capacity, $c_p/c_v$.  The speed of sound can be related to the adiabatic compressibility $\kappa_S$ and the isothermal compressibility $\kappa_T$ as follows: 
\begin{equation}
c = \left(\frac{1}{\rho \kappa_S}\right)^{1/2} = \left(\frac{c_p}{c_v}\frac{1}{\rho \kappa_T}\right)^{1/2},
\end{equation}
allowing us to re-write Bridgman's formula in terms of thermodynamic properties as
\begin{equation}
\lambda = 2.8k_{\textrm{B}}v^{-2/3}\left(\frac{1}{\rho \kappa_S}\right)^{1/2},
\end{equation}
and Eyring's formula as
\begin{equation}
\lambda = 2.8k_{\textrm{B}}v^{-2/3}\left(\frac{1}{\rho \kappa_T}\right)^{1/2},
\end{equation}
We have used the TSEOS \cite{Holten_2012b} to evaluate these expressions.

Bridgman's and Eyring's formulae differ little in the region where experimental data are available.  According to two-state thermodynamics, both the isothermal and the adiabatic compressibilities show maxima associated with the existence of a virtual liquid-liquid critical point.  These maxima are located close to the Widom line, defined as the locus of maximum fluctuations of the order parameter, a continuation of the liquid-liquid transition line into the one-phase region \cite{Fuentevilla_2006,Bertrand_2011,Holten_2012a,Holten_2012b}.  We note that the magnitudes of these maxima are strongly dependent on the value of the critical pressure, the value of which (possibly ranging from 10 to 40~MPa) is difficult to determine.  In conclusion, the observed behavior of thermal conductivity is anomalous inasmuch as it decreases upon cooling, it tracks the anomalous behavior of the adiabatic and isothermal compressibilities as shown in Fig. 3.

\begin{figure}[h!]
\includegraphics[width=0.49\textwidth]{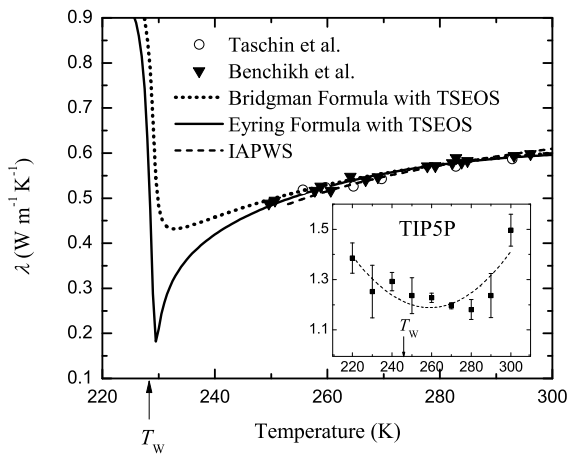}
\caption{Thermal conductivity calculated from thermal diffusivity \cite{Taschin_2006,Benchikh_1985} and the TSEOS \cite{Holten_2012b}.  The formulations due to Bridgman (4) and Eyring (5) are both shown, as is the IAPWS correlation for thermal conductivity (which is only guaranteed down to 273~K) \cite{IAPWS_ThermalConductivity_2012}.  The inset shows the simulation data of Kumar and Stanley for the TIP5P model \cite{Kumar_2011}, and the dashed curve on the inset is a quadratic fit to the simulation data.  $T_\textrm{W}$ refers to the Widom temperature in both the main graph and the inset.}
\end{figure}

Recently, Kumar and Stanley \cite{Kumar_2011} reported evidence of a thermal conductivity minimum in a simulation of the TIP5P model of water.  The simulation results indicated that the thermal conductivity of this model at first decreases upon cooling as has been observed experimentally in real water; it then reaches a minimum at approximately 255~K, and increases as the temperature is further decreased \cite{Kumar_2011}.  The latter increase might seem at first to conflict with experimental data for real water.  However, Kumar and Stanley's simulation locates the Widom temperature for the TIP5P model at roughly 245~K at atmospheric pressure, while thermodynamic equations of state place the Widom temperature for real water close to 228~K \cite{Fuentevilla_2006,Bertrand_2011,Holten_2012a,Holten_2012b}.  Rescaling the temperatures in the simulation results so that the Widom temperature occurs at 228~K places the predicted minimum at 237~K, several degrees below the lowest-temperature measurements of the thermal conductivity, so there is no real contradiction between simulation data and the predicted behavior of thermal conductivity in real water.  Moreover, one can expect the minimum of thermal conductivity observed in this model to be smoothed by finite-size effects, as is typical for simulations.  As can be seen in Fig. 3, both the Bridgman formula (4) and the Eyring formula (5) indicate that thermal conductivty should go through a minimum.  We shall return to this topic in more detail in Sec. IV.

Next, we address the possible effects of critical fluctuations on the thermal conductivity of supercooled water.  It is well documented that the thermal conductivity of water diverges at its vapor-liquid critical point, and the associated anomaly affects the thermal conductivity noticeably throughout the critical region \cite{Sengers_1985}.  We investigate the possibility of such a divergence of the thermal conductivity near the liquid-liquid critical point of H$_2$O, and any effects that this might have on the measurable behavior of the thermal conductivity in supercooled water.

\section{Predictions of Mode-Coupling Theory}
In the vicinity of a critical point, couplings among the various hydrodynamic modes of a system become increasingly significant as fluctuations in the system become long ranged.  This leads to anomalous behavior of the thermal conductivity in the critical region, including a divergence of the thermal conductivity at the vapor-liquid critical point.  This divergence has been observed in striking agreement with the mode-coupling theory in many molecular fluids near their respective vapor-liquid critical points \cite{Luettmer-Strathmann_1995, Sengers_1985}.  (This mode-coupling theory describes dynamics in the vicinity of a critical point and should not be confused with the mode-coupling theory of the glass transition).  Due to a mode-coupling contribution to the thermal diffusivity, which arises in molecular fluids from a coupling between the heat mode and the viscous mode, thermal conductivity can be expected to diverge near any critical point at which the isobaric heat capacity diverges more strongly than the correlation length $\xi$ \cite{Luettmer-Strathmann_1995}.  Such a strong divergence of the isobaric heat capacity is a feature of the TSEOS \cite{Holten_2012b}, as well as other related scaling models \cite{Bertrand_2011, Fuentevilla_2006,Holten_2012a}; this prompted us to investigate the possibility that a critical enhancement to the thermal conductivity might be experimentally observable.

In order to carry out our mode-coupling calculations we once again made use of the TSEOS \cite{Holten_2012b} for thermodynamic properties.  This formulation is renormalized by critical fluctuations, and asymptotically close to the critical point it is identical to the scaling models referred to above \cite{Bertrand_2011, Fuentevilla_2006,Holten_2012a}.  For the background value of the thermal conductivity we used the formulation of the International Association for the Properties of Water and Steam (IAPWS) \cite{IAPWS_ThermalConductivity_2012}.  IAPWS provides a formulation for the viscosity of water at atmospheric pressure that is valid for temperatures as low as 253.15~K \cite{IAPWS_Viscosity_2009,IAPWS_1Atm_2009}. Extrapolating below that value is a more subtle task.  Data taken by Osipov \emph{et al.} \cite{Osipov_1977} from 238~K to 273~K show a clear super-Arrhenius dependence on temperature (``fragile" behavior in Angell’s nomenclature \cite{Angell_1995}).  Some researchers \cite{Ito_1999} have found evidence of Arrhenius temperature dependence (``strong" behavior) close to the glass transition, and thus of a fragile-to-strong crossover in water; such a crossover has been observed to occur at 228~K in confined water \cite{Faraone_2004}.  Other measurements, however, have found super-Arrhenius behavior at the glass transition \cite{Smith_1999}.  Starr \emph{et al.} have used Adam-Gibbs theory to estimate the viscosity in the experimentally inaccessible region \cite{Starr_2003}, and this extrapolation includes a fragile-to-strong crossover as well. We have fit a super-Arrhenius equation to the experimental data of Osipov \emph{et al.} (see Fig. 4), and in making the fit we have chosen a hypothetical temperature of structural arrest so that our extrapolation agrees with that of Starr \emph{et al.} in the portion of the unstable region for which we make predictions.  In our calculations we use the IAPWS formulation for viscosity at atmospheric pressure  \cite{IAPWS_Viscosity_2009,IAPWS_1Atm_2009} for temperatures above 254~K and our extrapolation for temperatures below 254~K.

\begin{figure}[h!]
\includegraphics[width=0.49\textwidth]{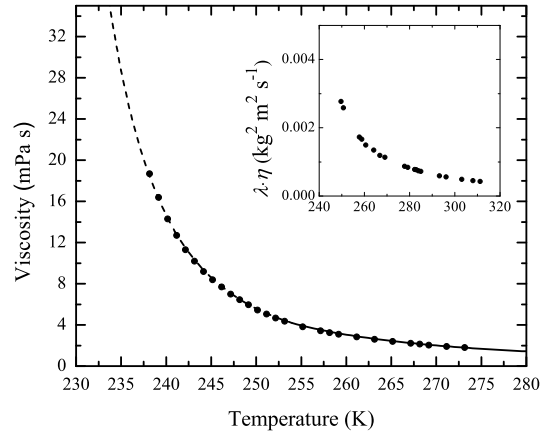}
\caption{Viscosity data taken by Osipov \emph{et al.} at atmospheric pressure (dots) \cite{Osipov_1977}.  The solid curve shows the IAPWS formulation for viscosity at atmospheric pressure \cite{IAPWS_Viscosity_2009,IAPWS_1Atm_2009}, while the dashed curve is a fit to the data of a super-Arrhenius or Vogel-Fulcher-Tamman law: $\eta = 0.0885 \exp{[220/(T-197)]}$, with $T$ in $K$.  The inset shows the product of viscosity and thermal conductivity.  The increase in the viscosity completely dominates the decrease in the thermal conductivity, so it is clear that the decrease in thermal conductivity is not a result of the increase in viscosity.}
\end{figure}

Mode-coupling theory gives a pair of coupled equations for the critical enhancements to viscosity and thermal diffusivity, and these equations should in principle be solved by iteration  \cite{Kawasaki_1970}. However, the viscosity anomaly associated with critical fluctuations is very weak, so we work only to one-loop order in the iteration and simply use the background value of viscosity, which is strongly temperature dependent (see Fig. 4).
The anomaly of the thermal diffusivity is additive in nature, meaning that the thermal diffusivity can be split into a background and a critical contribution \cite{Sengers_1985}:
\begin{equation}
a = a_\textrm{b} + \Delta a ,
\end{equation}
and the thermal conductivity can be treated similarly:
\begin{equation}
\lambda = \rho c_p a = \lambda_\textrm{b} + \Delta \lambda .
\end{equation}

Even to one-loop order the integral for the thermal diffusivity enhancement cannot be solved exactly except in the asymptotic limit $\xi \to \infty$.  Our approximation strategy for treating the crossover from critical to mean-field behavior is based on the model put forward by Olchowy and Sengers \cite{Olchowy_1989, Perkins_2013}.  It yields an expression of the form
\begin{equation}
\Delta\lambda = \rho c_p\Delta a = \rho c_p\frac{R_Dk_\textrm{B}T}{6\pi\eta\xi} \left(\Omega-\Omega_0\right),
\end{equation}
where $R_D$ is a universal amplitude very close to unity (experiments by Burstyn \emph{et al.} \cite{Burstyn_1980} find $R_D = 1.02 \pm 0.06$).
In the limit $\xi \to \infty$, we have $(\Omega - \Omega_0) \to 1$. If we adopt $R_D=1$, this expression tends to the well-known limit of a Stokes-Einstein law for thermal diffusivity in which the correlation length of the critical fluctuations replaces the hydrodynamic radius of Brownian particles:
\begin{equation}
\Delta a = \frac{k_\textrm{B}T}{{6\pi\eta\xi}}.
\end{equation}
Due to the effects of long-time tails on the hydrodynamic modes, mode-coupling effects do not completely vanish far from criticality \cite{Olchowy_1988,Olchowy_1989}.  These long-time effects are already present in the background, and so the phenomenological term $\Omega_0$ is introduced to subtract these effects from our expression so that it represents only the critical enhancement.  Further details of the approximation scheme can be found in the appendix to this article.

The path along atmospheric pressure is not the critical isobar, and along this path properties may exhibit finite anomalies but they do not diverge.  We find that at atmospheric pressure, the critical enhancement to the thermal conductivity would reach its maximum in the vicinity of the Widom temperature and at a value close to $\Delta \lambda = 8\times10^{-6}$ $\text{Wm}^{-1}\text{K}^{-1}$ (see Fig. 5).  This effect is certainly too small to be measurable, either in the experimentally accessible regime or in the region of the phase diagram below the homogeneous nucleation line (predictions for this region appear in Fig. 5 as a dashed curve).  Figure 5 shows experimental data as well as the prediction from Eyring's equation \cite{Hirschfelder_1954}.  Any effect from the critical enhancement is far too small to be visible on such a graph; it is shown in the inset, magnified by a factor of a million.  We note further that error bars on thermal-conductivity measurements in water are typically of the order of $10^{-2}$ $\text{Wm}^{-1}\text{K}^{-1}$, several orders of magnitude larger than any possible effect induced by critical fluctuations at atmospheric pressure.
Even at the critical pressure (for which the TSEOS uses 13.1 MPa \cite{Holten_2012b}), and even if one could somehow carry out measurements in ``no man's land", any critical enhancement would be undetectable, as it would be confined to to small a range of temperatures.  

\begin{figure}[h!]
\includegraphics[width=0.49\textwidth]{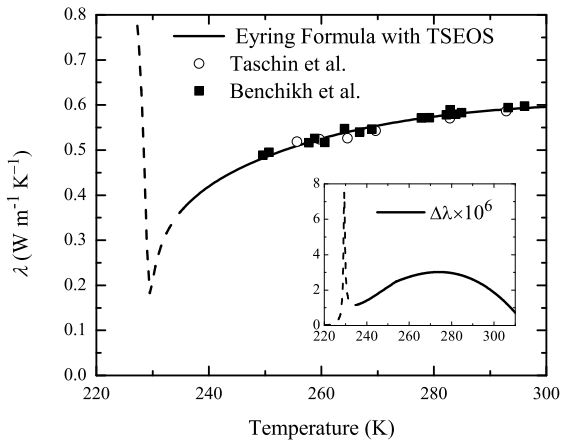}
\caption{The main figure shows thermal conductivity in water, both from experimental data \cite{Benchikh_1985,Taschin_2006} and from the Eyring's equation (5) evaluated with the TSEOS \cite{Holten_2012b}.  The inset shows the critical enhancement in the same units, but magnified by a factor of one million ($10^6$).  The curves change from solid to dashed at the temperature of homogeneous nucleation.  The critical effect is several orders of magnitude smaller than the background and will be completely undetectable.}
\end{figure}


\section{Discussion}
While thermal transport near the vapor-liquid critical point is dominated by a Stokes-Einstein law for thermal diffusivity, near the hypothesized liquid-liquid critical point thermal transport will continue to be governed by the thermodynamic properties.  We can identify two immediate reasons for this striking difference between the two critical regions.  First of all, mode-coupling theory predicts that the critical enhancement to the thermal conductivity will be inversely proportional to the viscosity.  The viscosity of supercooled water increases dramatically as the temperature decreases: Osipov \emph{et al.} \cite{Osipov_1977} found an increase of a factor of 10 between 273~K and 238~K in their experiment.  Our extrapolation predicts that the viscosity at the Widom temperature will be yet another order of magnitude larger than at 238 K. This means that the viscosity of water at the Widom temperature for the hypothesized second critical point is between two and three orders of magnitude larger than at the liquid-vapor critical point, which greatly suppresses any critical enhancement to the thermal conductivity.

The second physical reason why any measurable anomaly will be confined to such a tiny region of the phase diagram in supercooled water is that while temperature is (to a good approximation) the thermal field for the vapor-liquid phase transition, for the hypothesized liquid-liquid phase transition in water it very nearly plays the role of the ordering field $h_1$ \cite{Bertrand_2011, Fuentevilla_2006,Holten_2012a}.  Near the critical point, the ordering field is related to the order parameter according to a power law $\phi \sim h_1^{1/\delta}$, with $\delta \approx 4.8$ for the Ising-model universality class \cite{Landau_1980,Fisher_1983}.  Thus, asymptotically close to the critical temperature, small variations in the temperature correspond to large variations in the order parameter, and so for practical purposes a small deviation from the Widom temperature moves the system far from criticality.

The thermal-conductivity minimum merits some further discussion.  We have noted, following Benchikh \emph{et al.} \cite{Benchikh_1985}, that thermal conductivity in supercooled water follows the thermodynamic (low-frequency) speed of sound in good agreement with Bridgman (4) and Eyring's (5) formulae for thermal conductivity in a liquid.  Trinh \emph{et al.} have raised the possibility of a minimum in the speed of sound \cite{Trinh_1978}, and calculations from the TSEOS predict such a minimum at 233~K \cite{Holten_2012b}.  So it is plausible that thermal conductivity does indeed have a minimum, and that it continues to follow the speed of sound in the medium as it does throughout the measured part of the supercooled regime.  The anomalous behavior of the speed of sound probably results from the existence of a region of the phase diagram in which it is thermodynamically rather inexpensive to convert water between the HDL to LDL structures, leading to a higher compressibility.  Once this condition no longer obtains, more typical liquid behavior might be expected to resume.  In explaining their simulation results, Kumar and Stanley have noted that many other systems show an increase in thermal conductivity upon ordering  \cite{Kumar_2011}.  It is interesting to note that strong links between thermodynamics and dynamics are a general feature of the supercooled region of glass-forming liquids; a good discussion of the link can be found in \cite{Debenedetti_2001}.  However, the correlation between compressibility and thermal conductivity in water is robust in much of the stable liquid region as well and it is not associated with glassy dynamics. 

Finally, we note the possibility of an even more fundamental difference between water near its vapor-liquid critical point and water in the supercooled regime, one that points to a possible direction for further research.  In the vicinity of the vapor-liquid critical point, the fluctuations in the density are not associated with the conversion of molecules between distinct molecular structures.  No or little structural relaxation takes place, and the decay of density fluctuations is completely governed by diffusion dynamics. In supercooled water, contrarily, the existence of distinct molecular structures seems likely to introduce into the dynamics a relaxation time for conversion between these structures \cite{Mallamace_2013}.  Although liquid water at atmospheric pressure is non-dispersive down to temperatures of $-15$~\textcelsius , high-frequency sound ($\omega > 1~$GHz) below that temperature shows a positive dispersion relation.  This phenomenon has been widely attributed to a structural relaxation process taking place in supercooled water \cite{Cunsolo_1996,Magazu_1989,Taschin_2011}; in particular, it may be related to the conversion between the kinds of hydrogen bonding characteristic of HDL and LDL water.  We emphasize that while the dispersion of sound near the vapor-liquid critical point is solely an effect of the relaxation of critical fluctuations, the dispersion of sound in supercooled water is most likely a visco-elastic phenomenon \cite{Mallamace_2013}.  The characteristic time of this relaxation probably does not depend on the wavenumber of any fluctuation of the density, and such a relaxation will be present even in the mean-field approximation.  Near the liquid-liquid critical point, both kinds of effects---diffusive decay of density fluctuations and structural relaxation associated with hydrogen bonding---may be present.  Coupling between a relaxational visco-elastic mode and a diffusive decay mode has been observed and analyzed in polymers both near to a critical point and far from the critical region \cite{Kostko_2007}.  This analysis may provide a point of departure for further investigation of transport in supercooled water. 

\section{Conclusion}
At high temperatures, density fluctuations associated with the vapor-liquid critical point noticeably affect the thermal conductivity over a fairly broad temperature range.  Our calculations suggest, however, that this will not be the case for the hypothesized liquid-liquid critical point in supercooled water.   The effect of density fluctuations on the liquid-liquid critical point is too small to be measurable.  The Stokes-Einstein law that describes thermal transport in the vapor-liquid critical region will not be applicable to the thermal diffusivity in the liquid-liquid critical region.  On the other hand, the thermal conductivity and thermal diffusivity of supercooled water are strongly correlated with the anomalies of the thermodynamic properties associated with the existence of a liquid-liquid transition. The minimum of the thermal conductivity, found in simulations by Kumar and Stanley \cite{Kumar_2011}, should also exist in real water, being associated with the maximum of compressibility and with the minimum of the speed of sound.

Moreover, supercooled water may differ even more fundamentally from water in the vapor-liquid critical region due to the possible existence of a non-conserved order parameter associated with the structure of the water molecules \cite{Tanaka_1999}.  Rather than obeying the dynamics of diffusive decay, the relaxation of this order parameter would be associated with a characteristic time that is independent of wavenumber.  This would have far-reaching implications for various dynamic properties of supercooled water.

\acknowledgments
We are grateful to P. Kumar and H. E. Stanley for providing us with their simulation data on the TIP5P model and for useful comments.
Acknowledgment is made to the Donors of the American Chemical Society Petroleum Research Fund, for support of this research, Grant no. 52666-ND6.

\appendix*
\section{Details of the Mode Coupling Calculations}

We start from the mode-coupling integral as given in Ref. \cite{Perkins_2013}:
\begin{equation}
\Delta a = \frac{k_\textrm{B}T}{(2 \pi)^3} \int_0^{q_\textrm{D}} \textrm{d}\textbf{k}\frac{c_p(k)}{c_p(0)}\frac{k^{-2}\textrm{sin}^2(\theta )}{\eta(k)(1 + a(k)\rho /\eta(k))}.
\end{equation} 
The Olchowy-Sengers approximation \cite{Olchowy_1988} reasons that close to the critical point the term $a\rho/\eta$ is negligible.  As noted we use the background value of the viscosity in our approximation, so after carrying out the angular integrals the integral to evaluate is as follows:

\begin{equation}
\Delta a = \frac{k_\textrm{B}T}{3 \pi^2\eta}\int_0^{q_\textrm{D}} \textrm{d}k \frac{c_p(k)}{c_p(0)}.
\end{equation}

In the TSEOS the isobaric heat capacity can be expressed in terms of the relevant scaling variables and a few parameters of the system as follows:
\begin{equation}
c_p = -\lambda (1+b\Delta \hat{P}) \hat{T}(\phi + 1) + c_p^\textrm{A} + \frac{1}{2}\lambda ^2 (1+b\Delta \hat{P})^2 \hat{T}^2 \chi_1,
\end{equation}
where $\lambda$ and $b$ are parameters from the TSEOS with the values given in the supplement to ref. \cite{Holten_2012b}. $c_p^\textrm{A}$ is a background term representing the dimensionless heat capacity of pure HDL. The strongest divergence in the isobaric heat capacity is in the strong scaling susceptibility $\chi_1$, so for our third approximation we separate the term containing $\chi_1$ from the rest of the heat capacity, and treat the remaining terms (the sum of which we shall call $A$) as having no wave-number dependence.  We then use the Ornstein-Zernike approximation for the wave-number dependence of $\chi_1$:
\begin{equation}
\chi_1 (k) = \frac{\chi_1 (0)}{1 + k^2 \xi^2},
\end{equation}
where $\xi$ is the correlation length characterizing the fluctuations of the order parameter.

Henceforth $c_p$ without any explicit wave-number dependence will refer to the hydrodynamic value, $c_p(k \to 0)$.  With that notation, the integral that we must evaluate for the thermal diffusivity takes the form
\begin{equation}
\Delta a = \frac{k_\textrm{B}T}{3 \pi^2 \eta}\int_0^{q_\textrm{D}} \textrm{d}k \left[ \frac{c_p-A}{c_p}\frac{1}{1 + \xi^2 k^2}+\frac{A}{c_p} \right],
\end{equation}
which yields the modified Stokes-Einstein law
\begin{equation}
\Delta a = \frac{k_\textrm{B}T}{6\pi\eta\xi}\frac{2}{\pi}\left[\frac{c_p-A}{c_p}\textrm{arctan}(q_\textrm{D}\xi) + \frac{A}{c_p}q_\textrm{D}\xi\right].
\end{equation}
For convenience we define
\begin{equation}
\Omega = \frac{2}{\pi}\left[\frac{c_p-A}{c_p}\textrm{arctan}(q_\textrm{D}\xi) + \frac{A}{c_p}q_\textrm{D}\xi\right].
\end{equation}
This result is still not entirely satisfactory because in the limit of vanishing correlation length, that is, far from criticality, it does not vanish.  In fact:
\begin{equation}
\lim_{\xi \to 0} \frac{k_\textrm{B}T}{6\pi\eta\xi}\frac{2}{\pi}\left[\frac{c_p-A}{c_p}\textrm{arctan}(q_\textrm{D}\xi) + \frac{A}{c_p}q_\textrm{D}\xi\right] = \frac{k_\textrm{B}Tq_\textrm{D}}{3\pi^2\eta}.
\end{equation}
The physical reason for this as summarized by  Olchowy and Sengers in refs. \cite{Olchowy_1988,Olchowy_1989} is that mode coupling is also responsible for the ``long-time-tail effects on transport properties"\cite{Olchowy_1989}.  These effects are not critical effects and will be observed in the background, so if we want to find the effects due to critical fluctuations we should subtract off this remnant. For this reason we subtract the following term from $\Omega$:
\begin{equation}
\Omega_0(q_\textrm{D}\xi ) = \frac{2}{\pi} \left\lbrace 1-\textrm{exp}\left[ -\frac{q_\textrm{D}\xi}{1 + \left(1-A/c_p\right) (q\xi)^4} \right] \right\rbrace.
\end{equation}
This phenomenological expression has the following limiting behavior:
\begin{eqnarray}
\lim_{x\to 0} \Omega (x) = \frac{2}{\pi},
\\
\lim_{x\to \infty} \Omega (x) = 0,
\end{eqnarray}
so that we have 
\begin{equation}
\lim_{\xi \to 0} \Delta a = 0.
\end{equation}

The complete expression, using the above definitions, is
\begin{equation}
\Delta a = \frac{k_\textrm{B}T}{6\pi\eta\xi}\left(\Omega - \Omega_0\right).
\end{equation} 

Because we are interested only in the critical enhancement to the thermal conductivity, for the correlation length we should use only the critical enhancement to the correlation length.  We estimate the ``background" correlation length $\xi_{\textrm{b}}$ by observing the correlation length at a temperature $T_{\textrm{ref}}=315~K$ far from any critical point and assuming that the background correlation length in the system is proportional to the space between molecules.  So we have
\begin{equation}
\xi_{\textrm{b}}(T)=\frac{\xi(T_{\textrm{ref}})}{v(T_{\textrm{ref}})^{1/3}} v(T)^{1/3}.
\end{equation}
For the correlation length in our calculations we use
\begin{equation}
\xi_{\textrm{c}} = \xi - \xi_{\textrm{b}},
\end{equation}
where $\xi$ is the correlation length predicted by the TSEOS \cite{Holten_2012b}.

For the wave-number cutoff $q_\textrm{D}$, we used a correlation identified by Perkins \emph{et al.} between the wave-number cutoff and the amplitude $\xi_0$ of the correlation length anomaly \cite{Perkins_2013}:
\begin{equation}
q_\textrm{D}^{-1} = 3.683 \xi_0 - 1.336\times 10^{-10}~\textrm{m}.
\end{equation}

\bibliographystyle{apsrev}
\bibliography{../PhysicsRefs}

\end{document}